\def\bold#1{\setbox0=\hbox{$#1$}%
     \kern-.025em\copy0\kern-\wd0
     \kern.05em\copy0\kern-\wd0
     \kern-.025em\raise.0433em\box0 }
\def\slash#1{\setbox0=\hbox{$#1$}#1\hskip-\wd0\dimen0=5pt\advance
       \dimen0 by-\ht0\advance\dimen0 by\dp0\lower0.5\dimen0\hbox
         to\wd0{\hss\sl/\/\hss}}
\documentstyle[12pt]{article}

\newlength{\dinwidth}
\newlength{\dinmargin}
\newcommand{\resection}[1]{\setcounter{equation}{0}\section{#1}}

\begin{document}

\def\lq{\left [}
\def\rq{\right ]}
\def\LL{{\cal L}}
\def\VV{{\cal V}}
\def\AA{{\cal A}}
\def\MM{{\cal M}}
\def\qq{<{\overline u}u>}
\def\dmu{\partial_{\mu}}
\def\dmus{\partial^{\mu}}
\def\gi{{g_{P^* P \pi}}}
\def\gid{{g_{D^* D \pi}}}
\def\gib{{g_{B^* B \pi}}}

\newcommand{\be}{\begin{equation}}
\newcommand{\ee}{\end{equation}}
\newcommand{\bea}{\begin{eqnarray}}
\newcommand{\eea}{\end{eqnarray}}
\newcommand{\nn}{\nonumber}
\newcommand{\dd}{\displaystyle}

\thispagestyle{empty}
\vspace*{4cm}
\begin{center}
  \begin{Large}
  \begin{bf}
CHIRAL EFFECTIVE THEORY FOR HEAVY MESONS
\\
  \end{bf}
  \end{Large}
  \vspace{5mm}
  \begin{large}
G. Nardulli\\
  \end{large}
Dipartimento di Fisica, Univ.
di Bari\\
I.N.F.N., Sezione di Bari\\
  \vspace{5mm}
\end{center}
  \vspace{2cm}
\begin{quotation}
\vspace*{1.5cm}
\begin{center}
  \begin{bf}
  ABSTRACT
  \end{bf}
\end{center}
\vspace*{0.5cm}
I review recent developments in the description of the interactions
between light and heavy mesons by an effective chiral lagrangian
having the symmetries of the Heavy Quark Effective Theory. In
particular the problem of the determination
of the strong coupling constants $\gi$
and $g_{P^* P \rho}$ ($P, P^*=$ heavy mesons) is addressed.
\noindent
\end{quotation}
  \vspace{35mm}
\begin{center}
BARI-TH/94-188\\\
November 1994\\\
\end{center}
\vspace{1cm}
Invited talk given at the 18th J. Hopkins Workshop (Florence, September 1994)
\noindent
\newpage

\setcounter{page}{1}
\resection{Introduction}

A recent and interesting development of the Heavy Quark Effective Theory (HQET)
\cite{isgur} is represented by the study of the chiral effective
lagrangian containing heavy mesons \cite{wise}, \cite{alii}, \cite{cinesi},
\cite{light2}. The idea is to generalize chiral effective theory
by terms containing heavy mesons, described by effective field
operators, as well as their interactions with the light mesons.

A consistent way to implement this program should not only define the heavy
meson field operators, but also respect all the symmetries of HQET, i.e.,

a) spin symmetry; a well known consequence of it is the fact that
heavy mesons can be organized in spin multiplets, e. g. the $0^-$ and the $1^-$
mesons;

b) heavy flavour symmetry; this symmetry holds provided $m_Q>>\Lambda_{QCD}$
and applies to quantities that remain finite in the limit $m_Q \to \infty$;

c) velocity superselection rule, which implies that the effective lagrangian
describing strong interactions
should be written as a sum of terms that are diagonal in the
velocity dependent heavy meson field operators.

Moreover the lagrangian should respect chiral symmetry not only in the
terms containing light fields, but also in the part
containing the heavy meson operators.

An important step of this program is the determination of
the coupling constants of the lagrangian. To this issue I shall devote
most of this talk. This determination can render chiral effective theory
for heavy mesons a useful tool to deal with different aspects of the
interactions between light and heavy mesons as well as with their weak
and electromagnetic interactions.

\resection{The chiral effective lagrangian for light and heavy mesons}

To begin with, we describe the effective field operators appearing in
the chiral lagrangian.
Negative parity heavy $Q{\bar q}_a$ mesons are represented by field
operators in the form of a $4 \times 4$ Dirac matrix
\bea
H_a &=& \frac{(1+\slash v)}{2}[P_{a\mu}^*\gamma^\mu-P_a\gamma_5]\\
{\bar H}_a &=& \gamma_0 H_a^\dagger\gamma_0
\eea
Here $v$ is the heavy meson velocity, $a=1,2,3$
(for $u,d$ and $s$ respectively),
$P^{*\mu}_a$ and $P_a$ are annihilation operators normalized as follows
\bea
\langle 0|P_a| Q{\bar q}_a (0^-)\rangle & =&\sqrt{M_H}\\
\langle 0|P^*_a| Q{\bar q}_a (1^-)\rangle & = & \epsilon^{\mu}\sqrt{M_H}
\eea
with $v^\mu P^*_{a\mu}=0$ and $M_H=M_P=M_{P^*}$, is the
heavy meson mass.
The pseudoscalar light mesons are described, as usual in chiral
effective theory, by
\be
\xi=\exp{\frac{iM}{f_{\pi}}}
\ee
where
\be
{M}=
\left (\begin{array}{ccc}
\sqrt{\frac{1}{2}}\pi^0+\sqrt{\frac{1}{6}}\eta & \pi^+ & K^+\nn\\
\pi^- & -\sqrt{\frac{1}{2}}\pi^0+\sqrt{\frac{1}{6}}\eta & K^0\\
K^- & {\bar K}^0 &-\sqrt{\frac{2}{3}}\eta
\end{array}\right )
\ee
where $f_{\pi} \simeq 132 MeV$ is the pion leptonic decay
constant in the chiral limit. Under the chiral symmetry the fields transform as
follows
\bea
\xi & \to  & g_L\xi U^\dagger=U\xi g_R^\dagger\\
\Sigma & \to  & g_L\Sigma {g_R}^\dagger\\
H & \to  & H U^\dagger\\
{\bar H} & \to & U {\bar H}
\eea
where  $\Sigma=\xi^2$, $g_L$, $g_R$ are global $SU(3)$
transformations and $U$ is a
function of $x$, of the fields and of $g_L$, $g_R$.

The lagrangian describing the fields $H$ and $\xi$ and their interactions,
under the hypothesis of chiral and spin-flavour symmetry and at the lowest
order in light mesons derivatives is \cite{wise},\cite{alii}, \cite{cinesi}
\cite{light2} :
\be
\LL_{0}=\frac{f_{\pi}^2}{8}<\partial^\mu\Sigma\partial_\mu
\Sigma^\dagger > +i < H_b v^\mu D_{\mu ba} {\bar H}_a >
+i g <H_b \gamma_\mu \gamma_5 \AA^\mu_{ba} {\bar H}_a>
\ee
where $<\ldots >$ means the trace, and
\bea
D_{\mu ba}&=&\delta_{ba}\partial_\mu+\VV_{\mu ba}
=\delta_{ba}\partial_\mu+\frac{1}{2}\left(\xi^\dagger\partial_\mu \xi
+\xi\partial_\mu \xi^\dagger\right)_{ba}\\
\AA_{\mu ba}&=&\frac{1}{2}\left(\xi^\dagger\partial_\mu \xi-\xi
\partial_\mu \xi^\dagger\right)_{ba}
\eea
Besides chiral symmetry, which is obvious, since, under chiral
transformations,
\bea
D_\mu {\bar H} \to U D_\mu {\bar H} \nn\\
\AA_\mu \to U \AA_\mu U^\dagger \; ,
\eea
the lagrangian (2.11) possesses the heavy quark spin symmetry $SU(2)_v$, which
acts as
\bea
H_a &\to& {\hat S} H_a \\
{\bar H}_a &\to& {\bar H}_a {\hat S}^\dagger
\eea
with ${\hat S}{\hat S}^\dagger=1$ and $[\slash v,{\hat S}]=0$, and a heavy
quark flavour symmetry arising from the absence of terms containing
$m_Q$.

Explicit symmetry breaking terms can  also be introduced, by
adding to $\LL_0$
an  extra piece $\LL_1$ containing corrections
at the lowest order in $m_q$ and $1/m_Q$. I do not write down it here
\cite{wise}, \cite{light2} but it is worth stressing that it is
precisely this term which is responsible for the light pseudoscalar
meson masses and for the  mass difference $\delta m_H$
between the particles $P$ and $P^*$ contained in the field $H$, i.e.:
\be
 M_P=M_H ~~~~~~~~~~ M_{P^*}=M_H+\delta m_H
\ee

The vector meson resonances belonging to the low-lying $SU(3)$ octet can be
introduced by using the hidden gauge symmetry approach
\cite{light1}, \cite{light2}
(for another approach see \cite{Schec}).
The new lagrangian containing these particles, to be added to $\LL_0+
\LL_1$, is as follows:
\bea
\LL_2&=& -\frac{f^2_{\pi}}{2}a <(\VV_\mu-
\rho_\mu)^2>+\frac{1}{2g_V^2}<F_{\mu\nu}(\rho)F^{\mu\nu}(\rho)> \nn\\
&+&i\beta <H_bv^\mu\left(\VV_\mu-\rho_\mu\right)_{ba}{\bar H}_a>\nn\\
&+&\frac{\beta^2}{2f^2_{\pi} a}<{\bar H}_b H_a{\bar H}_a H_b>+
i \lambda <H_b \sigma^{\mu\nu} F_{\mu\nu}(\rho)_{ba} {\bar H}_a>
\eea
where $F_{\mu\nu}(\rho)=\partial_\mu\rho_\nu-\partial_\nu\rho_\mu+
[\rho_\mu,\rho_\nu]$, and $\rho_\mu$ is defined as
\be
\rho_\mu=i\frac{g_V}{\sqrt{2}}\hat\rho_\mu
\ee
$\hat\rho$ is a hermitian $3\times 3$ matrix analogous to (2.6) containing
the light vector mesons $\rho^{0,\pm}$, $K^*$, $\omega_8$. $g_V$, $\beta$,
$\lambda$
and $a$ are coupling constants; by imposing the two KSRF relations
\cite{light2} one obtains
\be
a=2 \ \ \ \ \ \ \ \ \ \ \ \ \ \ \ g_V \approx 5.8
\ee

The resulting effective lagrangian
$\LL=\, \LL_0 \, + \, \LL_1 \, + \, \LL_2$ can be generalized to include
low-lying positive parity $Q{\bar q}_a$ heavy meson states.
For $p$ waves ($l=1$), the heavy quark effective theory predicts two
distinct multiplets, one containing a $0^+$ and a $1^+$ degenerate states,
the other one comprising a $1^+$ and a $2^+$ state \cite {iw}, \cite {Ros}
\cite {Falk}.
Their inclusion is needed if one wishes
to describe heavy mesons
semileptonic decays into a final state containing a light
vector meson. We refer to the literature for such an analysis \cite{light2}.
In the remaining part of this paper I will address the problem of the
determination of the unknown strong coupling constants $g$ and $\lambda$
(no determination of $\beta$ is available yet).

\resection{Semileptonic decays}

I now wish to show that semileptonic decays of the
heavy pseudoscalar mesons into final states containing
 light mesons can be used to determine the strong coupling constants
$g$ and $\lambda$. In doing that, however one has to make a further
assumption concerning the $q^2$ behaviour of the semileptonic form
factors.

At the lowest order in derivatives of the pseudoscalar couplings and in the
symmetry limit, weak interactions between light pseudoscalars and a
heavy meson are described
by the weak current \cite {wise}:
\be
L^{\mu}_a=\frac{i {\hat F}}{2}
<\gamma^{\mu} (1-\gamma_5) H_b \xi^{\dagger}_{ba}>
\ee
where ${\hat F}$ is related to the pseudoscalar heavy meson decay constant
$f_P$,
defined by
\be
<0|\overline{q}_a \gamma^{\mu} \gamma_5 Q|P_b (p)>=ip^{\mu} f_P \delta_{ab}
\ee
as follows:
\be
{\hat F}=f_P \sqrt{M_H} \; .
\ee
${\hat F}$ is finite in the $m_Q \to \infty$ limit and is independent
of the heavy quark mass, but for logarithmic corrections
that are however tiny. Its numerical value can be inferred by
QCD sum rules analyses \cite{neubert}:
\be
{\hat F}= 0.41 \pm 0.04 \; GeV^{3/2}
\ee
This result is obtained including radiative $O(\alpha_s)$ corrections.

Let us first consider the semileptonic decay of the heavy
pseudoscalar meson into a light peudoscalar meson. To be definite
we consider the decay:
\be
B \to \pi \ell \nu_{\ell}
\ee
The hadronic matrix element
can be written in terms of the form factors $F_0$, $F_1$ as follows
\be
<\pi(p')|V^{\mu}|B(p)> =
 \big[ (p+p')^{\mu}+\frac{M_\pi^2-M_B^2}{q^2} q^{\mu}\big]
F_1(q^2) -\frac {M_\pi^2-M_B^2}{q^2} q^{\mu} F_0(q^2)
\ee
where $q^{\mu}=(p-p')^{\mu}$, $F_0(0)=F_1(0)$.
The form factors $F_0$ and $F_1$
take contributions, in a dispersion relation, from the $0^+$ and $1^-$ meson
states respectively.

Using the chiral lagrangian and the current (3.1),
one obtains,
at the leading order in $1/m_Q$ and at $q^2=q^2_{\rm max}$,
the following result \be
F_1(q^2_{\rm max})=\frac {g {\sqrt {M_B}} {\hat F}} {2 f_{\pi}
(v\cdot k -\delta m_H)}
\ee
whereas $F_0(q^2_{\rm max})$ is found to satisfy the analogous of the
Callan Treiman relation in the chiral limit \cite{light2}. It is worth
stressing that this result
arises from a polar diagram with a $B^*$ exchange;
$k^{\mu}$ is the residual momentum
related to the physical momenta by $k^{\mu}=q^{\mu}-M_{B^*} v^{\mu}$ (and
$p^{\mu}=M_B v^{\mu}$).

A similar analysis can be performed for the semileptonic decay process with
a light vector in the final state. For definiteness we examine
\be
B \to \rho \ell \nu_{\ell}
\ee
and we consider only the vector current matrix element
\be
<\rho (\epsilon,p')| V^{\mu}|B(p)> =
\frac {2 V(q^2)} {M_B+M_{\rho}}
\epsilon^{\mu \nu \alpha \beta}\epsilon^*_{\nu} p_{\alpha} p'_{\beta}
\ee

In a dispersion relation the form factor $V(q^2)$ takes contribution from
the $1^-$
pole, i.e. the $B^*$ particle.

Using the chiral lagrangian and the current (3.1)
one gets,
at $q^2=q^2_{\rm max}$ and at leading order in $1/m_Q$ the result
\be
V(q^2_{\rm max})=-\frac {g_V} {\sqrt 2} \lambda {\hat F}
\frac {M_B+M_{\rho}} {{\sqrt {M_B}} (v\cdot
k -\delta m_B)}
\ee

These results are obtained in the chiral limit
and for
$m_Q \to \infty$; in these limits they apply not only to the
decays $B \to \pi \ell \nu_{\ell}$ or $B \to \rho \ell \nu_{\ell}$,
but also, using
the simmetries of the effective
current and lagrangian, to
e.g. $D \to \pi \ell \nu_{\ell}$,
$D \to \rho \ell \nu_{\ell}$,
$D \to K \ell \nu_{\ell}$,
$D \to K^* \ell \nu_{\ell}$ etc.
Therefore we could use experimental data on these decays to fix
the unknown quantities $g$ and $\lambda$.
In order to make  contact with the experimental data, however, we have to
make an ansatz on the $q^2$
behaviour of the form factors.
The contributions we have written down arise from polar diagrams,
which suggests a simple pole behaviour. This is also hinted by
the QCD sum rules analysis contained in Ref. \cite{ball}.
Therefore we
assume for the form factors $F_1(q^2)$ and  $V(q^2)$
 the generic formula
\be
F(q^2)= \frac {F(0)} {1 - \frac {q^2}{m^2}}
\ee
For the pole masses we use the masses
indicated by a pole dominated dispersion relation \cite {Korner}.

For the $D \to \pi$ semileptonic decay one gets:
\bea
F_1(0)&=& -\frac{g {\hat F}}{2 f_{\pi}}
\sqrt{M_D} \;\frac {M_{D^*}+M_D -M_{\pi}}
{M_{D^*}^2} \; .
\eea
{}From the experimental
value $|F_1(0)|=0.79\pm 0.20$, one gets:
\be
|g|=0.40\pm 0.10 \label{cc} \ee

This result agrees, within the errors,
 with the result obtained using as an input
$D\to K$ semileptonic decay \cite{Stone}.

Let us now turn to semileptonic decays into vector mesons. The experimental
input we can use from $D \to K^* \ell \nu_{\ell}$ is as follows:
\be
V(0) = 0.95\pm 0.20 \label{VV}
\ee
This is an average among the data
from the different Fermilab experiments \cite {E653}, \cite
{E691}. More recent data from the E-687 Collaboration \cite{E687}
agree,
 within the errors, with \ref{VV}.  The calculated weak coupling at $q^2=0$ is:
\bea
V(0)& = & \frac {g_V \lambda}{\sqrt 2} \frac {(M_D+M_{K^*}) (M_{D^*}
+M_D-M_{K^*})}{M_{D^*}^2} \frac {\hat F}{\sqrt{M_D}}
\eea
and one numerically obtains
for $\lambda$:

\be |\lambda|=0.40 \pm 0.10 \; GeV^{-1} \hskip 3 pt , \label{cd}
\ee
It is interesting to compute, by the values of $g$ and
$\lambda$ in
Eqs. (\ref{cc}), (\ref{cd}) and by using previous results for
$F_1(0)$ and $V(0)$ (adapted to $B$ case) the values of the form factors
for
the $B \to \pi \ell \nu_{\ell}$ and $B \to \rho \ell \nu_{\ell}$
decays. One obtains
\be
|F_1^{B \to \pi}(0)| \; =\; 0.53 \pm 0.13
\ee
\be
|V^{B \to \rho}(0)| \; = \; 0.61 \pm 0.15 \; .
\ee
To conclude,
Eqs. (\ref{cc}), (\ref{cd}) are the results
for the strong coupling constants of the chiral effective
lagrangian obtained by the analysis of
semileptonic decays. Since they are based on an extra assumption
(polar $q^2$ behaviour of the form factors $F_1$ and $V$) and
on the neglect of the heavy mass corrections, it is worth looking for
different theoretical and/or phenomenological determinations.

\resection{Radiative decays}

In this section we shall show
that the results of Eqs. (\ref{cc}) and (\ref{cd})
are compatible with the experimental data on the radiative
decay
\be D^* \to D \gamma \hskip 3 pt . \label{eq : 18}
\ee
The matrix element for this decay can be written as follows:

\be \MM(D^* \to D \gamma) =e \; \epsilon^{* \mu} J_\mu \hskip 3 pt
\label{eq : 19} \ee

\noindent with:
\bea J_\mu &=& <D(p^{\prime})|J_\mu^{em}|D^*(p, \eta)>=
\nonumber \\
&=& <D(p^{\prime})|e_Q \bar{Q} \gamma^\mu Q + e_q \bar{q} \gamma^\mu q|D^*(
p, \eta)>= \label{eq : 20} \\
&=& e_Q J_\mu^Q + e_q J_\mu^q \nonumber \hskip 3 pt , \eea

\noindent where $e_Q={2 \over 3}$ is the heavy quark ($Q=c$) charge and $e_q$
is the light quark charge ($e_q=e_u=2/3$ for $D^{*0}$ and $e_q=-1/3$ for
$D^{*+}$ and $D^*_s$). Let us consider the two currents appearing in
(\ref{eq : 20})
separately. $J_\mu^Q$ can be expressed in terms of the Isgur-Wise
universal form factor \cite{isgur} as follows:

\be <D(p^{\prime})|\bar{c} \gamma^\mu c|D^*(p, \eta)>=
    i  \sqrt{M_D M_{D^*}} \xi(v \cdot v^{\prime})
\epsilon^{\mu \nu \alpha \beta}
\eta_\nu v_\alpha v^{\prime}_\beta  \hskip 3 pt , \label{eq : 21} \ee

\noindent where $p^{\prime}=M_D v^{\prime}$, $p=M_{D^*} v$ and $v
\cdot v^{\prime} \simeq 1$ at
$q^2=0$.

As for vector current containing light quarks $J^q_\mu$, one can assume
Vector Meson Dominance \cite {defazio1}
and
write:

\be J^q_\mu= \sum_{V, \lambda}  <D(p^{\prime}) V(q, {\epsilon_1}(\lambda))|
D^*(p, \eta)> { i \over  q^2-M_V^2}
<0|\bar{q} \gamma_\mu q|V(q, {\epsilon_1}(\lambda))>
 \label{eq : 25} \ee

\noindent where $q^2=0$ and the sum is over the vector meson resonances $V=
\omega
$, $\rho^0$, $\phi$ and over the $V$ helicities.
The vacuum-to-meson current matrix
element appearing in (\ref{eq : 25})
is given by:

\be <0|\bar{q} T^i \gamma^\mu q|V(q, { \epsilon_1})>
={\epsilon_1}^\mu f_V \;
 \hskip 3 pt , \label{eq : 26} \ee

\noindent From $\omega \to e^+ e^-$ and $\rho^0 \to e^+ e^-$ decays
\cite{pdg} we get
$f_{\rho} = f_{\omega} = 0.17 \hskip 3 pt GeV^2$; from
$\phi \to e^+ e^-$ we have $f_{\phi}=0.25 GeV^2$.
Using (\ref{eq : 26}) and the
strong lagrangian $\LL_2$ we can easily compute $J^q_\mu$ and therefore
(\ref{eq : 20}). The results are \cite{defazio1}:

\be \MM(D^* \to D \gamma)=i \; \epsilon^{\mu \nu \alpha \beta}
\epsilon^*_\mu \eta_\nu v_\alpha v^{\prime}_\beta \sqrt{M_D M_{D^*}}[e_Q-e_q 2
\sqrt{2} g_V \lambda M_{D^*} {f_V \over M^2_{\omega}}] \hskip 3 pt ,
\label{eq : 27} \ee

\be \MM(D^*_s \to D_s \gamma)=i \; \epsilon^{\mu \nu \alpha \beta}
\epsilon^*_\mu \eta_\nu v_\alpha v^{\prime}_\beta \sqrt{M_{D_s} m_{D^*_s}}
[e_Q+{1 \over 3} 2 \sqrt{2} g_V \lambda M_{D^*_s} {f_{\phi} \over M^2_{\phi}}]
\hskip 3 pt ,\label{eq : 28} \ee

\noindent where $e_Q=e_c={2 \over 3}$.
Eq.(\ref{eq : 27}) holds for both $D^{*+} \to
D^+ \gamma$ and  $D^{*0} \to D^0 \gamma$ (with $e_q=-{1 \over 3}$ and ${2 \over
3}$ respectively). We can now use
the determination of $\lambda$ contained in the previous section
to obtain the radiative widths. Since we have only obtained
the absolute value of $\lambda$,
we have to fix the sign, which can be done by imposing that
the relative sign between the two contributions is identical to the one given
by the constituent quark model \cite{eichten}, i.e. we take
\be
\lambda=-0.40 \pm 0.10 \; .
\ee
It is clear that Eqs.(\ref{eq : 27}), (\ref{eq : 28}) describe with
obvious changes also $B^*$ radiative decays.
\par
{}From the amplitudes (\ref{eq : 27}), (\ref{eq : 28})
we can compute radiative decay rates for $D^*$ and
$B^*$.
Moreover one can compute
the decay width for the process
\be
D^* \to D \pi \hskip 3 pt , \label{dp}
\ee
that can be written in terms of the matrix element
\be
<\pi^-(q)~D^o(q_2) | D^{*-}(q_1,\epsilon )> \; = \; \gid \;
\epsilon^{\mu} \cdot q_{\mu}
\label{sda}
\ee
The strong coupling constant $\gid$ is related to the scaled
constant $g$ of the effective chiral
lagrangian
by the formula
\be
\gid={{2 M_D}\over{f_{\pi}}} g \; . \label{ginf}
\ee
which is valid in the infinite heavy quark mass limit.

The numerical results for the   $D^*$ and $B^*$  decay widths
are reported in Table I together with the CLEO data
\cite{cleo} on  radiative $D^*$ decays.
We observe an overall
agreement between theoretical results and experiment,
which we interpret as a
corroboration of the numerical values indicated by the
semileptonic decays. We point out however that if we use semileptonic $D$
decays to predict radiative and strong $D$ decays we do not actually
test the heavy flavour symmetry and, in particular, the
results for $g$ and $\lambda$ that have been obtained
could be effective values, containing a heavy quark mass dependence. This is
the reason to get independent determinations of these constants, as
we will see in the next sections.

It can be finally observed that
the general
structure of the matrix elements (\ref{eq : 27}) and (\ref{eq : 28})
of \cite {defazio1} coincides with
analyses of
other authors \cite{eichten}, \cite{eletsky}, \cite{amundson}, \cite{georgi},
\cite{cheng},
but there are some numerical differences mainly
arising from the light quark current that
is not provided by the heavy quark effective theory and
is treated
by different authors in different manners.
  \vspace{10mm}
\begin{table}
\begin{center}
\begin{tabular}{l c c }
 & {\bf Table I} &  \\ & & \\
 \hline \hline
Decay rate/ BR & theory & experiment \\ \hline
$\Gamma(D^{*+})$ & $46.1 \pm 14.2 \hskip 3 pt KeV$ &  $<131 \hskip 3 pt KeV \;$
\cite{accmor}
\\ \hline
$ BR(D^{*+} \to D^+ \pi^0)$ & $31.2 \pm 17.4 \%$ & $30.8 \pm 0.4 \pm 0.8$ \\
\hline
$BR(D^{*+} \to D^0 \pi^+)$ & $67.7 \pm 34.2 \%$ & $68.1 \pm 1.0 \pm 1.3$ \\
\hline
$BR(D^{*+} \to D^+ \gamma)$ & $1.1 \pm 0.9 \%$ & $1.1 \pm 1.4 \pm 1.6$ \\
\hline \\ \hline
$\Gamma(D^{*0})$ & $36.7 \pm 9.7 \hskip 3 pt KeV$ &  \\ \hline
$ BR(D^{*0} \to D^0 \pi^0)$ & $56.4 \pm 27.1 \%$ & $63.6 \pm 2.3 \pm 3.3$ \\
\hline
$BR(D^{*0} \to D^0 \gamma)$ & $43.6 \pm 17.8 \%$ & $36.4 \pm 2.3 \pm 3.3$ \\
\hline  \\ \hline
$\Gamma(D^*_s)=
\Gamma(D^*_s \to D_s \gamma)$ & $(0.24 \pm 0.24) \hskip 3  pt KeV$ &  \\ \hline
\\ \hline
$\Gamma(B^{*+})=\Gamma(B^{*+} \to B^+ \gamma)$ & $(0.22 \pm 0.09)
\hskip 3  pt KeV$ &  \\ \hline \\  \hline
$\Gamma(B^{*0})=\Gamma(B^{*0} \to B^0 \gamma)$ & $(0.075 \pm 0.027)
\hskip 3  pt KeV$ &  \\ \hline \hline

\end{tabular}
\end{center}
\end{table}
\vspace{20 mm}
\vspace{20 mm}

\resection{Quark model determination of the strong coupling
constant $g$}

The previous analyses, based on semileptonic and radiative decays
of heavy mesons point to a rather small value of the strong coupling constant
 $g$ appearing in the effective chiral lagrangian. In the literature
one can find the value $g \simeq 1$, \cite{cheng} as given by the
non relativistic potential model. It is interesting to
show that the quark model result for $g$ can be reconciled with
our previous finding $g \simeq 0.40$ provided one
takes into account the relativistic motion of the light
quark inside the
heavy meson \cite{defazio2}. This analysis is based
on a QCD inspired relativistic potential model
where the heavy hadrons
$D_a$ and $D^*_a$, made up by the quark $Q$ and the antiquark ${\bar q_a}$, are
described by  a relativistic wavefunction
$\psi(\vec{k} + x\vec{p} , -\vec{k} + (1-x) \vec{p})$.
This wavefunction satisfies the
Salpeter equation \cite{salpeter}
\bea
\Big\{ \sqrt{(\vec{k}+x \vec{p})^2 + m_Q^2} &+&
\sqrt{[-\vec{k}+(1-x) \vec{p}]^2 + m_{q_a}^2}-\sqrt{M_D^2+\vec{p}^2} \Big\}
\psi( \vec{k} + x \vec{p}, -\vec{k} + (1-x) \vec{p}) \nonumber \\
&+& \int d \vec {k^{\prime}} V(\vec{p}, \vec{k}, \vec{k^{\prime}})
  \psi (\vec{k^{\prime}}, \vec{p}-\vec{k^{\prime}})=0 \label{eq : 7}
\eea
that arises from
the bound-state Bethe Salpeter equation by considering the instantaneous time
approximation and restricting the Fock space to the $Q{\bar q}$ pairs (for
more
details see \cite{pietroni}). The
Salpeter equation includes relativistic effects due to the
kinematics explicitly and is valid in a moving frame where the meson $D$
(or $D^*$),
having mass $M_D$, has momentum $\vec{p}$; the wave function $\psi$ is
normalized as follows:

\be {1 \over (2\pi)^3} \int d \vec{k} |\psi|^2=2\sqrt{M_D^2+\vec{p}^2}
\hskip 3 pt , \label{eq : 8} \ee
Note that the quark $Q$
and the antiquark ${\bar q_a}$, carry momenta $\vec{k} + x \vec{p}$ and
$-\vec{k} + (1-x) \vec{p}$ respectively.

The instantaneous potential $V$ coincides, in the meson rest
frame, with the Richardson potential \cite{richardson};
in the $r$-space
it grows linearly when $r \to \infty$ and follows QCD
predictions for small $r$.
 In order
to avoid unphysical singularities \cite{nardulli}, one
assumes that $V(r)$, near
the origin, is constant:

\be V(r)=V(r_M) \hskip 1 cm \left( r \leq r_M={\lambda'  \over 3 M_D }{4 \pi
\over 3} \right)
\hskip 3 pt . \label{eq : 10} \ee

\noindent The values of the parameters, as obtained by fits to meson masses,
are as follows:
$m_u=m_d=38 \hskip 3 pt MeV$; $m_s=115 \hskip 3 pt  MeV$, $m_c=1452
\hskip 3 pt MeV$, $m_b=4890 \hskip 3 pt MeV$,
$\lambda'=0.6$.

In order to compute the strong coupling constant
$g$
one expresses
the axial current $A_\mu$ containing the light quarks in terms of
quark operators. Taking the derivative of $A_\mu$,
one obtains $(J_5=i{\bar d} \gamma_5 u)$:

\be (m_u+m_d)<D^0(k)|J_5|D^{*+}(p,\epsilon)>=-i(\epsilon \cdot q)\; 2M_{D^*}
A_0(q^2) \hskip 3 pt . \label{eq : 14} \ee

\noindent where $A_0(q^2)$ is a form
factor that, for small $q^2$,
is dominated by the $\pi$ pole. One therefore  obtains, for $q^2$ small:

\be g_{D^* D \pi} ={M_\pi^2 -q^2 \over M_\pi^2 } {2 M_{D^*} \over f_\pi} A_0(
q^2) \hskip 3 pt , \label{eq : 15} \ee

\noindent which shows that, in the chiral limit
$(q^2=0)$,

\be g =  A_0(0) \hskip 3 pt .
\label{a0} \ee

\noindent If
 $E_q=\sqrt{k^2 +m_q^2}$, $m_u=m_d=m_q$ and $\tilde{u}(
k)$ is related to the wave function $\psi$ with
$\vec{p}=0$ by the equation:

\be \tilde{u}(k)={k \; \psi(k) \over \sqrt{2} \pi } \hskip 3 pt ,
\label{ut}
\ee
one obtains

\be g=A_0(0)=
{1 \over 4 M_D} \int_0^\infty dk |\tilde{u}(k)|^2 {E_q + m_q
\over E_q} \left[ 1-{k^2 \over 3 (E_q+m_q)^2} \right] \hskip 3 pt .
\label{equaz}
\ee

It is interesting to consider
immediately the non-relativistic limit, where: $E_q \simeq
m_q \gg k$. In this limit one obtains:

\be g={1  \over 2 M_D} \int_0^\infty dk |\tilde{u}(k)|^2=1  \label{cin}
\ee

\noindent because of the normalization of the wavefunction. Eq.
(\ref{cin}) reproduces the well known constituent quark model result
\cite{cinesi},\cite{isgur}.\par Let us now take in (\ref{equaz}) the limit
$m_q \to 0$, which is possible since we work in the chiral limit and there is
no restriction to the values of $m_q$ in the Salpeter equation
 In this case, we obtain:

\be g={1 \over 3} \hskip 3 pt . \label{chir} \ee

\noindent It is worth to stress that the strong reduction of the value of $g$
from the naive non relativistic quark constituent model value ($g=1$) to
the
result (\ref{chir}) has a simple explanation in the effect of the
relativistic kinematics taken into account by the Salpeter equation.

If one introduces light quark masses
as given by
the fit of the meson masses ($m_q=38$
$MeV$ \cite{pietroni}), one has to consider
Eq. (\ref{equaz}).
$\tilde{u}(k)$ is obtained by solving the
Salpeter equation numerically
by the Multhopp method \cite{multhopp}.
In this case one obtains
\be
g\simeq 0.39 \; ,
\ee a result in agreement
with the previous  determination
 based on the semileptonic $D$ decays.

\resection{QCD sum rule calculation of $g$}

Finally I report on a recent calculation based on
QCD sum rules \cite{qcdsr} (for other similar calculations see
 \cite{eletsky},
\cite{grozin}).

Let us consider the off-shell process
\be
B^{*-}(\epsilon,q_1) \to {\bar B^0} (q_2) + \pi^-(q) \; .
\ee
One considers the
correlator
\be
A_{\mu}(P,q) = i \int dx <\pi^-(q)| T(V_{\mu}(x) j_5(0) |0> e^{-iq_1x} = A
q_{\mu} + B P_{\mu}
\label{corr}
\ee
where $V_{\mu}={\overline u} \gamma_{\mu} b$, $j_5=i{\overline b} \gamma_5 d$,
$P=q_1+q_2$ and $A$, $B$ are scalar functions of $q_1^2$, $q_2^2$, $q^2$.

Both $A$ and $B$ satisfy dispersion relations and are computed, according to
the QCD sum rules method, in two ways: either by saturating the dispersion
relation by physical hadronic states or by means of the operator product
expansion (OPE). Considering
the invariant function $A$
in the soft pion limit ($q \to 0$) and for large Euclidean
momenta ($q_1^2=q_2^2 \to - \infty$)
and performing the OPE one has the following result
\be
A=A^{(0)}+A^{(1)}+A^{(2)}+A^{(3)}+A^{(4)}+A^{(5)}
\label{terms}
\ee
with
\bea
A^{(0)}&=&{{-1}\over{q_1^2-m_b^2}} \lq m_b f_{\pi} +{{\qq}\over{f_{\pi}}}\rq
\nn\\
A^{(1)}&=&-{2\over 3} {{1}\over{q_1^2-m_b^2}} {{\qq}\over{f_{\pi}}} \lq
{{m_b^2}\over{q_1^2-m_b^2}} -2 \rq \nn\\
A^{(2)}&=&{{m_b f_{\pi} m_1^2}\over{9 (q_1^2-m_b^2)^2}} \lq 1+ {{10
m_b^2}\over{q_1^2- m_b^2}} \rq - {{m_o^2 \qq}\over{4 f_{\pi} (q_1^2-m_b^2)^2}}
\lq 1- {{2 m_b^2}\over{q_1^2 -m_b^2}} \rq \nn\\
A^{(3)}&=&{{m_0^2 \qq}\over{6 f_{\pi}}} \lq {1\over{(q_1^2-m_b^2)^2}} -{{2
m_b^2}\over{(q_1^2-m_b^2)^3}} +{{6 m_b^4}\over{(q_1^2-m_b^2)^4}} \rq \nn\\
A^{(4)}&=&{1\over{(q_1^2-m_b^2)^2}} \lq {{m_0^2 \qq}\over{4 f_{\pi}}} +m_b
f_{\pi} m_1^2 \rq \nn\\
A^{(5)}&=&{{m_0^2 \qq}\over{6 f_{\pi}}} \lq {1\over{(q_1^2-m_b^2)^2}} -{{2
m_b^2}\over{(q_1^2-m_b^2)^3}} \rq \; .
\label{cntr}
\eea
In eqs.(\ref{cntr}) $\qq$ is the quark condensate ($\qq =-(240 MeV)^3$), $m_0$
and $m_1$ are defined by the equations
\be
<{\overline u} g_s \sigma \cdot G u> =m_0^2\qq
\ee
and
\be
<\pi (q)|{\overline u} D^2 \gamma_{\mu} \gamma_5 d |0> = -i f_{\pi} m_1^2
q_{\mu}
\ee
and their numerical values are: $m_0^2=0.8 \; GeV^2$, $m_1^2=0.2 \; GeV^2$
\cite{novikov,chernyak}.

We now write down the hadronic side of the sum rule.
In the dispersion relation
\be
A(0,q_1^2,q_2^2)={1\over{\pi^2}} \int ds ds' {{\rho(s,s')}\over{(s-q_1^2)
(s'-q_2^2)}} \; .
\label{intg}
\ee
one  divides the integration region into three parts. The first region (I)
is the
square given by $ m_b^2 \le s \le s_0$, $m_b^2 \le s'\le s_0$;
for $s_0$ small enough, (I) contains
only the $B$ and $B^*$ poles, whose contribution is
\bea
A_I(0,q_1^2,q_2^2)&=&{{f_B f_{B^*} M_B^2}\over{4 m_b M_{B^*}}} \;
\Big[ {{\gib (3
M_{B^*}^2+M_B^2)}\over{(q_1^2-M_{B^*}^2) (q_2^2-M_B^2)}} +\nn\\
&+& {{\gib}\over{q_1^2-M_{B^*}^2}}+{{3 f_+ - f_-}\over{q_2^2 -M_B^2}} \Big]
\label{uno}
\eea
where $f_B$, $f_{B^*}$, are the usual leptonic decay
constants, while $f_+$ and $f_-$ are defined by
\be
<\pi^-(q) {\overline B}^0(q_2) | B^{*-}(q_1)> = (f_+ P_{\mu} - f_- q_{\mu})
\epsilon^{\mu} \; .
\ee

The remaining integration regions in (\ref{intg}) contain new
unknown couplings. However one can get rid of them
as well as of the term containing $3 f_+ - f_-$ in (\ref{uno}). Indeed the
unwanted terms (the so called "parasitic terms")
for $q_1^2=q_2^2$ are proportional, after the Borel transformation, to
$1/M^2$, while the contribution we are interested in,
i.e. the term containing
the factor $\gib (3
M_{B^*}^2+M_B^2)$, after the Borel transform, gives rise to a
contribution proportional to
$1/M^4$, if one neglects the tiny mass difference between
$M_{B^*}$ and $M_B$. Therefore one can exploit the different $M^2$ behaviour to
isolate the relevant contribution \cite{qcdsr}. Without going into details,
I report here the result of this analysis for the case $m_Q \to \infty$.

The infinite heavy quark mass limit ($m_b \to \infty$)
is performed according to the usual procedure \cite{shuryak},
\cite{neubert} \cite{paver}.
In terms of low energy parameters the quantities appearing in the
finite mass sum rule are written as follows:
\bea
M_B &=& m_b+\omega\nn\\
M_{B^*}-M_B &=& {\cal{O}} \left( {1\over{m_b}}\right)\nn\\
f_B &=& f_{B^*} = {{\hat F}\over{\sqrt {m_b}}} \;
\label{par}
\eea
\noindent $\omega$ represents the binding energy of the meson, which is finite
in the limit $m_b \to \infty$;
$\hat F$ has been computed by QCD sum rules:
for $\omega=0.625$ GeV and the threshold
$y_0 = {s_0 - m_b^2 \over 2 m_b}$ in the range $1.1 - 1.4 \;  GeV$
the result is
 ${\hat F}= 0.30 \pm 0.05 \; GeV^{3/2}$ (at the order $\alpha_s=0$)
and  ${\hat F}= 0.41 \pm 0.04 \; GeV^{3/2}$
(including radiative corrections), as we have stressed already. \cite{neubert}.

The sum rule for $g$ is derived after
having expressed the Borel parameter $M^2$ in terms of the low energy parameter
$E$: $M^2=2 m_b E$. One readily obtains:
\bea
g &=& {{f_{\pi}^2}\over{{\hat F}^2}} e^{\omega /E}\Big\{{1\over{y_0-\omega}}
\Big[ \omega^2 \left( 1 -{\qq\over{3f_{\pi}^2 E}}-{{5 m_1^2}\over{36 E^2}} +
{{m_0^2\qq}\over{48 E^3 f_{\pi}^2}} \right) + \nn\\
&-& 2\omega \left({\qq\over{3f_{\pi}^2}}+{{5 m_1^2}\over{18 E}} -
{{m_0^2\qq}\over{16 E^2 f_{\pi}^2}} \right) -{{5 m_1^2}\over{18}}+{{m_0^2\qq}
\over{8 E f_{\pi}^2}}\Big]+\nn\\
&+& \omega \left( 1 -{\qq\over{3f_{\pi}^2 E}}-{{5 m_1^2}\over{36 E^2}} +
{{m_0^2\qq}\over{48 E^3 f_{\pi}^2}} \right)- {\qq\over{3f_{\pi}^2}}+\nn\\
&-& {{5 m_1^2}\over{18 E}} +{{m_0^2\qq}\over{16 E^2 f_{\pi}^2}} \Big\}
\label{sr2}
\eea
We observe that the sum rule only gives the combination
${\hat F}^2 \; g$; therefore the result has a strong dependence on
${\hat F}$ .
This sum rule must be studied in the region of the external parameter $E$ where
the OPE is assumed to converge and where the contribution of higher resonances
is small ("duality" region); moreover the various terms of the OPE
should display a hierarchical structure, according to their dimension.
The corresponding result is, without inclusion of $O(\alpha_s)$ corrections:
\be g = 0.44 \pm 0.10 GeV^3 \; .
\label{r1}
\ee
One may have a hint on the possible role of the
$O(\alpha_s)$ corrections, by considering only
those induced by $\hat F$, that are available and should
represent the largest part of such corrections \cite{qcdsr};
in this case one would get
\be g \simeq 0.24 \label{r2}
\ee
\noindent The difference between (\ref{r1}) and (\ref{r2}) reflects
the well known important role of radiative corrections in the determination of
$f_B$ by QCD sum rules in the $m_Q \to \infty$ limit \cite{broad}. Results
compatible, within the theoretical uncertainties, with \ref{r1} have been
obtained by light cone sum rules in \cite{br}.

\resection{Conclusions}

The strong couplings $g$ and $\lambda$ play an important role in heavy meson
phenomenology. They are relevant in the strong
and radiative decays of the heavy mesons
and are expected to be important for their semileptonic decays
into final states containing light mesons.
They are also important inputs in the effective
chiral lagrangians for heavy mesons.
I have presented several ways
to determine these constants: by semileptonic decays \cite{light2},
using analysis
of radiative transitions \cite{defazio1}, a relativistic potential model
approach \cite{defazio2} and QCD sum rules \cite{qcdsr}.
These results can be summarized as follows:
\be
\lambda=-0.40 \pm 0.10
\ee
\be
|g|=0.25 \; - \; 0.50 \; ,
\ee
where the interval of values for $|g|$ represent a realistic range
of values as derived from previous analyses.

\par\noindent
{\bf Acknowledgements}

It is a pleasure to thank R. Casalbuoni, P. Colangelo, A. Deandrea,
F. De Fazio, N. Di Bartolomeo, F.Feruglio and R. Gatto for their
collaboration on the work reported in this paper.


\newpage

\begin{thebibliography}{99}

\bibitem{isgur}
N.Isgur and M.B.Wise, Phys. Lett. {\bf B232} (1989) 113; ibidem
{\bf B237} (1990) 527; M.B.Voloshin and M.A.Shifman, Sov.J.Nucl.Phys. {\bf
45} (1987) 292; ibidem {\bf 47} (1988) 511;
 H.D.Politzer and M.B. Wise, Phys. Lett.
{\bf 206B} (1988) 681; ibidem {\bf 208B} (1988) 504;
 E.Eichten and B.Hill, Phys.
Lett. {\bf 234B} (1990) 511; H.Georgi, Phys.Lett. {\bf 240B} (1990) 447;
B.Grinstein, Nucl. Phys. {\bf B339} (1990) 253; A.F.Falk, H.Georgi, B.Grinstein
and M.B.Wise, Nucl. Phys. {\bf B343} (1990) 1.

\bibitem{wise}
M.B.Wise, Phys. Rev. {\bf D45} (1992) R2188.


\bibitem{alii}
G.Burdman and J.F.Donoghue, Phys. Lett. {\bf B280} (1992) 287;
P.Cho Harvard preprint HUTP-92/A014 (1992).


\bibitem{cinesi}
T.-M. Yan, H.-Y. Cheng, C.-Y. Cheung, G.-L. Lin, Y. C. Lin and H.-L. Yu, Phys.
Rev. {\bf D46} (1992) 1148.


\bibitem{light2}
R.Casalbuoni, A.Deandrea, N.Di Bartolomeo, R.Gatto, F.Feruglio and G.Nardulli,
Phys. Lett. {\bf B299} (1993) 139.

\bibitem{light1}
R.Casalbuoni, A.Deandrea, N.Di Bartolomeo, R.Gatto, F.Feruglio and G.Nardulli,
Phys. Lett. {\bf B292} (1992) 371.

\bibitem{Schec}
J.Schechter and A.Subbaraman, Preprint SU-4240-519, September 1992.

\bibitem{iw}
N.Isgur and M.B.Wise, Phys. Rev. Lett. {\bf 66} (1991) 1130; Phys. Rev. {\bf
D43} (1991) 651.

\bibitem{Ros}
J.Rosner, Comm. Nucl. Part. Phys. {\bf 16} (1986) 109.

\bibitem{Falk}
A.F.Falk and M.Luke, Phys. Lett {\bf B292} (1992) 119; U.Kilian, J.C. K\"orner
and D. Pirjol, Phys. Lett. {\bf B288} (1992) 360.

\bibitem{neubert} M.Neubert, Phys. Rev. {\bf D 45} (1992) 2451.

\bibitem{ball}
P.Ball, Phys. Rev. {\bf D 48} (1993) 3190.

\bibitem{Korner}
J.G.K\"orner, K.Schilcher, M.Wirbel and Y.L.Wu,
Zeit. Phys. {\bf C48} (1990) 663.

\bibitem{Stone}
S.Stone, Syracuse University Report No. HEPSY-1-92.

\bibitem{E653}
K.Kodama et al.,  E-653 Collaboration,
Phys. Rev. Lett. {\bf 66} (1991)1819; Phys. Lett. {\bf B263}
(1991) 573; Phys. Lett. {\bf B286} (1992) 187.

\bibitem{E691}
J.C.Anjos et al., E-691 Collaboration,
Phys. Rev. Lett. {\bf 62} (1989) 1587; Phys. Rev. Lett. {\bf
65} (1990) 2630; Phys. Rev. Lett. {\bf 67} (1991) 1507.

\bibitem{E687}
P. L. Frabetti et al., E-687 Collaboration,
Phys. Lett. {\bf B 307} (1993) 262.

\bibitem{defazio1} P. Colangelo, F. De Fazio and G. Nardulli,
Phys. Lett. {\bf B 316 } (1993) 555.

\bibitem{pdg}
Particle Data Group, Review of Particle Properties, Phys. Rev. {\bf D45} (1992)
S1.

\bibitem{eichten}
E. Eichten, K. Gottfried, T. Kinoshita, K. D. Lane and T. M. Yan, Phys. Rev.
{\bf D21} (1980) 203.

\bibitem{cleo}
F. Butler et al., CLEO Collaboration Phys. Rev. Lett. {\bf 69} (1992) 2041.

\bibitem{eletsky}
V. L. Eletsky and Ya. I. Kogan, Zeit. fur Phys. {\bf C 28} (1985) 155.


\bibitem{amundson}
J. F. Amundson, C. G. Boyd, E. Jenkins, M.Luke, A. V. Manohar, J. L. Rosner, M.
J. Savage and M. B. Wise, Phys. Lett. {\bf B296} (1992) 415.

\bibitem{georgi}
P.Cho and H.Georgi, Phys. Lett. {\bf B296} (1992) 408;  Phys. Lett.
{\bf B300} (1993) 410 (E).

\bibitem{cheng}
H.-Y. Cheng, C.-Y. Cheung, G.-L. Lin, Y. C. Lin, T.-M. Yan, H.-L. Yu, Phys.
Rev. {\bf D47} (1993) 1030.


\bibitem{accmor}
The ACCMOR collaboration (S. Barlag et al.), Phys.Lett. {\bf B278} (1992) 480.

\bibitem{defazio2} P. Colangelo, F. De Fazio and G. Nardulli,
Phys. Lett. {\bf B 334 } (1994) 175.

\bibitem{salpeter}
E. E. Salpeter, Phys. Rev. {\bf 87} (1952) 328.

\bibitem{pietroni}
P. Colangelo, G. Nardulli and M. Pietroni, Phys. Rev. {\bf D43} (1991) 3002.

\bibitem{richardson}
J. L. Richardson, Phys. Lett. {\bf B 82} (1979) 272.

\bibitem{nardulli}
P. Cea and G. Nardulli, Phys. Rev. {\bf D34} (1986) 1863.


\bibitem{isgur}
N. Isgur and M. B. Wise, Phys. Rev. {\bf D41} (1990) 151.

\bibitem{multhopp}
K. Karamcheti, Principles of ideal fluid aerodynamics (Wiley, New York, 1966).


\bibitem{qcdsr}
P. Colangelo, G.
Nardulli, A. Deandrea, N. Di Bartolomeo, R. Gatto and
F. Feruglio, Phys. Lett. {\bf B 339 } (1994) 151.

\bibitem{grozin}
A.G.Grozin and O.I.Yakovlev, preprint BUDKERINP-94-3 (hep-ph/9401267).

\bibitem{novikov}
V.A.Novikov, M.A.Shifman, A.I.Vainshtein, M.B.Voloshin and V.I.Zakharov, Nucl.
Phys. {\bf B 237} (1984) 525.

\bibitem{chernyak}
A.R.Zhitnitskii, I.R.Zhitnitskii and V.L.Chernyak, Sov. J. Nucl. Phys. {\bf 38}
(1983) 645.

\bibitem{shuryak}
E.Shuryak, Nucl. Phys. {\bf B 198} (1982) 83.


\bibitem{paver}
P.Colangelo, G.Nardulli and N.Paver,
Proceedings of the ECFA Workshop on a European B-Meson Factory, R.Aleksan and
A.Ali {\it Eds.}, ECFA 93/151, pag.155.



\bibitem{broad}
D.J.Broadhurst and A.G.Grozin, Phys. Lett. {\bf B 274} (1992) 421.

\bibitem{br}
V. M. Belyaev, V. M. Braun, A. Khodjamirian and R. R\"uckl, preprint
MPI-PhT/94-62,  CEBAF-TH-94-22, LMU 15/94 (September 1994).

\end{thebibliography}
\end{document}